\documentclass[twocolumn,10pt]{IEEEtran}

\input epsf
\usepackage{graphicx}  
\usepackage{graphicx,epstopdf}   
\usepackage{pifont}
\usepackage{bm}
\usepackage{subcaption}
\usepackage{cite}
\usepackage{float}
\usepackage{amsfonts,balance}
\usepackage[utf8x]{inputenc}
\usepackage{cite}
\usepackage{color}
\usepackage{multirow}

\usepackage[linesnumbered,algoruled,boxed,lined]{algorithm2e} 

\IEEEoverridecommandlockouts
\usepackage{mathtools}
\usepackage{amsthm}
\usepackage{amsmath}
\newtheorem{theorem}{Theorem}

\newtheorem{corollary}{Corollary}

\usepackage{bm,cite,algorithmic,float,amsmath,amssymb}

\usepackage{dsfont}
\usepackage{amsmath}
\usepackage{cite}
\usepackage{citesort}
\usepackage{balance}
\usepackage[utf8x]{inputenc}

\IEEEoverridecommandlockouts

\usepackage{graphicx,epstopdf}
\usepackage{epsfig}	
\usepackage{amsfonts,balance}
\usepackage{bbm}

\usepackage{xcolor,colortbl}
\definecolor{Gray}{gray}{0.9}

\usepackage[
top    = 1.70cm,
bottom = 1.05in,
left   = 0.60 in,
right  = 0.60 in]{geometry}

\begin{document}


\title{Traffic Prediction and Random Access Control Optimization: Learning and Non-learning based Approaches}

\author{
\IEEEauthorblockN{Nan Jiang, \emph{Student Member, IEEE}, Yansha Deng, \emph{Member, IEEE}, and Arumugam Nallanathan, \emph{Fellow, IEEE}}



\thanks{This work was supported by the Engineering and Physical Sciences Research Council (EPSRC), U.K., under Grant EP/R006466/1. (Corresponding author: Yansha Deng.)}
\thanks{N. Jiang, and A. Nallanathan are with the School of Electronic Engineering and Computer Science, Queen Mary University of London, London E1 4NS, UK (e-mail: \{nan.jiang, a.nallanathan\}@qmul.ac.uk).}
\thanks{Y. Deng is with the Department of Informatics, King's College London, London WC2R 2LS, UK (e-mail: yansha.deng@kcl.ac.uk).}


\vspace*{-0.5cm}
}



\maketitle

\vspace*{-0.3cm}

\begin{abstract}
Random access schemes in modern wireless communications are generally based on the framed-ALOHA (f-ALOHA), which can be optimized by flexibly organizing devices' transmission and re-transmission. However, this optimization is generally intractable due to the lack of information about complex traffic generation statistics and the occurrence of the random collision. In this article, we first summarize the general structure of access control optimization for different random access schemes, and then review the existing access control optimization based on Machine Learning (ML) and non-ML techniques. We demonstrate that the ML-based methods can better optimize the access control problem compared with non-ML based methods, due to their capability in solving high complexity long-term optimization problem and learning experiential knowledge from reality. To further improve the random access performance, we propose two-step learning optimizers for access control optimization, which individually execute the traffic prediction and the access control configuration. In detail, our traffic prediction method relies on online supervised learning adopting Recurrent Neural Networks (RNNs) that can accurately capture traffic statistics over consecutive frames, and the access control configuration can use either a non-ML based controller or a cooperatively trained Deep Reinforcement Learning (DRL) based controller depending on the complexity of different random access schemes. Numerical results show that the proposed two-step cooperative learning optimizer considerably outperforms the conventional Deep Q-Network (DQN) in terms of higher training efficiency and better access performance.

\end{abstract}
\vspace*{-0.1cm}

\vspace*{-0.3cm}
\section{Introduction}
\vspace*{-0.0cm}

To achieve effective radio access, the random access technique has been integrated into multiple access protocol as a key component of modern wireless communication systems, e.g. Long-Term Evolution (LTE, a.k.a., 4G), Fifth Generation New Radio (5G NR) systems, and etc.. Taking 4G/5G cellular networks as an example, the random access technique is adopted by Random Access CHannel (RACH) procedure, which is used to establish or re-establish connection between an unsynchronized device and its associated Base Station (BS) \cite{LTE2013dahlman}. The reason to adopt random access is due to its minimum requirements of priori information, where devices randomly select channels and transmit preambles/packets to the associated BS without negotiation. This uncoordinated transmission inevitably brings uncertainty such that multiple devices may select the same channel at the same time, which results in collided signals that generally cannot be decoded by the BS. Severe collisions occur when massive devices simultaneously access, which results in access delay, packet loss, or even service unavailability. With the growing number of devices, such as in massive Internet of Things (mIoT), massive access by shared radio channels may create the heavy network overload problem, which brings one of the key challenges in communication networks, and motivates us to concentrate on massive random access in this article.
\color{black}


The random access framework provides the flexibility of designing access schemes to organize devices' transmission and re-transmission. For instance, Access Class Barring (ACB) Scheme can defer devices to transmit preamble according to a probability in order to alleviate network congestion. To better manage access, access control parameters (e.g., ACB factor), are expected to be flexibly selected according to the communication environment and traffic statistics. However, such flexible selection of the access control parameters in RACH is general intractable, due to the lack of knowledge at the BS regarding future traffic and channel statistics. 
\color{black}
To solve this problem, classical works \cite{wu2012fast,he2015traffic,duan2016d,sharma2019collaborative,jian2013novel,lin2014prada,vazquez2017contention} have devoted substantial efforts in designing efficient access control optimization techniques by deriving explicit optimization solutions based on a mathematical model that captures the regularities of the practical communication environment (to be detailed in Sec. \ref{sec4a}). However, their access performances are generally limited, due to the high complexity of the problem and the fact that the physics-based model is hard to accurately describe the communication environment.

\captionsetup{singlelinecheck=true}
\begin{table*}[htbp!]
	\centering
	\caption{RACH Protocols and Relevant Optimizations \vspace*{-0.0cm}}
	{\renewcommand{\arraystretch}{0.6}
		\begin{tabular}{|l|l|l|l|l|l|l|}
			\hline
			\rowcolor{Gray}
			\multicolumn{7}{|l|}{ $\vphantom{\Big(}$\bf{Comparison of RACH Protocols}}
			\\ \hline
		   \multicolumn{2}{|l|}{\bf{Solution}} &    \bf{KPI}  &     \multicolumn{2}{|l|}{\bf{Parameters} } &  \bf{Exact Control}  &   \bf{Reference}$\vphantom{\Big(}$  \\  \hline
		   \multicolumn{2}{|l|}{$\vphantom{\big(}$Access Class Barring (ACB)} & Success Accesses & 
		   \multicolumn{2}{|l|}{ACB factor} &  \checkmark  &                      \cite{wu2012fast,he2015traffic,duan2016d,Luis2018Reinforcement} \\
		   \multicolumn{2}{|l|}{$\vphantom{\big(}$Dynamic Resource Allocation (DRA)} & Time Delay & 
		   \multicolumn{2}{|l|}{Channels} &  \checkmark  & \cite{sharma2019collaborative} \\
		   \multicolumn{2}{|l|}{$\vphantom{\big(}$Back-Off (BO)} & Success Accesses & 
		   \multicolumn{2}{|l|}{BO factor} &  $\times$  & \cite{jian2013novel}\\
           \multicolumn{2}{|l|}{$\vphantom{\big(}$Prioritized Access} & Success Accesses & 
           \multicolumn{2}{|l|}{Access Periodicity} &  $\times$  & \cite{lin2014prada}\\
           \multicolumn{2}{|l|}{$\vphantom{\big(}$Distributed Queuing (DQ)} & Success Accesses  & 
           \multicolumn{2}{|l|}{Depth and breadth of the tree} &  $\times$  & \cite{vazquez2017contention}\\
            \hline
            
			\hline
			
			\rowcolor{Gray}
		 \multicolumn{7}{|l|}{ $\vphantom{\Big(}$\bf{Comparison of Optimization Methods}}
		\\  \hline
		   \bf{Type}  &    \bf{Sub-type}  &    \bf{Performance}  &     \bf{Complexity}  &  \bf{Online Adaptation}  &   $\vphantom{\Big(}$\bf{Training Efficiency}  & \bf{Reference} \\  \hline
		   $\vphantom{\big(}$\multirow{3}{*}{Non-ML based} & DA optimizer & Low &  Low &   $\times$   &  \textbf{-}     &                \cite{wu2012fast} \\
		    $\vphantom{\big(}$ & MoM optimizer & Low & Low &   $\times$  &  \textbf{-}  & \cite{duan2016d} \\
		    $\vphantom{\big(}$ & MLE optimizer & Moderate &  Moderate &  $\times$ &  \textbf{-}  & \cite{he2015traffic}\\
		    \hline
            $\vphantom{\big(}$\multirow{3}{*}{ML based}& RL-based optimizer  & High &  High &  \checkmark &  Slow & \cite{Luis2018Reinforcement}\\
           $\vphantom{\big(}$ &  SL-based optimizer$^*$ & High (Limited) & Moderate &  \checkmark &  Fast &  \cite{jiang2019online}\\
           $\vphantom{\big(}$ &  CPCL-based optimizer  & High &  High &  \checkmark &  Fast &  This work \\
                        \hline
		\end{tabular}
	}
	\label{table2}
	\\{\hspace*{+0.8cm} \raggedright $\vphantom{\big(}$ $*$ \scriptsize The SL-based optimizer only offers high performance in the RACH schemes with exact configuration solutions, e.g., ACB and the resource allocation schemes. \par}
  	\vspace*{-0.1cm}
\end{table*}

In this article, we first briefly introduce the RACH procedure and related RACH schemes in cellular-based networks. After that, we propose the fundamental mechanism of access control process, and conclude the state-of-the-art conventional dynamic access control techniques. Finally, we elaborate that Machine Learning (ML) based access control optimization has potential to better optimize the random access KPIs, due to its capability in addressing high complexity problem and learning experiential knowledge from environment.
\color{black}
Specifically, we introduce the-state-of-arts model-free Reinforcement Learning (RL) based access control optimization \cite{Luis2018Reinforcement,jiang2019deep}, which provides one-step solution of both the traffic prediction and the access control configuration. With the obtained optimal solution only relying on interacting with network environment, this one-step RL-based method requires minimal domain knowledge of the communication model, while it also suffers from low training efficiency and huge computational resource consumption. To solve these problems, we then propose a novel two-step learning framework for access control optimization by decomposing the learning process into two independent processes, which are the traffic prediction and the access control configurations, respectively. The traffic prediction is based on the online Supervised Learning (SL) adopting Recurrent Neural Networks (RNNs) given in \cite{jiang2019online}, and the access control configuration based on either a non-ML based controller or a cooperatively trained RL based controller. This two-step learning framework is proposed based on the fact that the access control configuration strongly depends on the forthcoming traffic load, and the design of decomposition between prediction and control considerably improves its training efficiency.

The remainder of the article is organized as follows. Section II illustrates the structure and research challenges of random access. Section III discusses the background of access control optimization and reviews existing non-ML based access control optimization methods. Section IV proposes learning-based access control optimization, including one-step RL-based method, SL-based traffic prediction, and the integrated SL-based prediction and RL-based configuration. Finally, Section V summarizes the conclusion and future work.
\color{black}

\section{Research Challenges and Random Access Schemes}

The RACH procedure is initialized by transmitting a preamble along with three control signals transmitted via scheduled channels. Generally, a device performs the \emph{contention-based} access via transmitting a randomly selected preamble to initiate RACH, in which the contention refers to that the preamble can be erroneously decoded due to collision, i.e., two or more devices transmit using the same preamble at the same time. To overcome such channel resource under-provision, several efficient random access schemes has been developed over the last decade. In the following, we first introduce the framework and research challenges of RACH in the cellular networks, and then describe the classical RACH schemes.

\subsection{RACH framework and research challenges}

The contention-based access requires multiple steps of control information exchanges between a device and its associated BS, which are usually handled via two different strategies: \emph{(a)} multi-step \emph{grant-based} procedure used in the conventional cellular networks, such as LTE and 5G NR; and \emph{(b)} two-step \emph{grant-free} procedure, which is proposed to handle the sporadic traffic of low latency IoT networks. In the grant-based RACH, the preamble transmission at step 1 is in the random access manner, whereas the steps 2-4 use dedicated channels scheduled by the BS that only occurs when the preamble transmission succeeds. This successive execution inevitably increase latency, but improves the system reliability. In the grant-free access, the preamble, the control information, and the data are integrated into a single sequence to be transmitted to its associated BS without any negotiation, which decreases the RACH delay with the sacrifice of the system reliability.

The contention-based RACH is built on the f-ALOHA structure, where each device is informed about the pool of available channels, and selects one of the channel uniformly at random for transmission. In particular, the size of channel pool can be flexibly defined by the BS, and the notification occurs at the beginning of each frame via broadcasting. Recent works \cite{wu2012fast,he2015traffic,duan2016d,sharma2019collaborative,jian2013novel,lin2014prada,vazquez2017contention} on f-ALOHA networks have focused on designing effective random access schemes and the optimization techniques for the purpose of handling access overload. To evaluate the the performance of the novel techniques, a list of KPIs are presented as follows:

\begin{itemize}
\item Access Success Probability (Reliability): a statistical probability mapping devices to complete access within a limited number of frames.
\item Access Delay: the time elapsed from the start of access to the time receiving the confirmation of access success.
\item Energy Consumption: the total energy consumed during RACH, which is mainly affected by the re-access times. 

\end{itemize}

\subsection{Random Access Schemes}\label{sec3a}

To support massive and diverse access requirements, existing literature have proposed RACH solutions in various wireless networks, including, but not limit to, grant-based RACH in LTE, 5G NR, NarrowBand IoT (NB-IoT), and etc.. These solutions are based on the f-ALOHA framework, and share the same purpose of providing more efficient access by alleviating the collisions during RACH. In general, the key idea of these solutions aim at overcome the channel resource under-provision by intelligently organizing devices' transmission and re-transmission. A classification of existing schemes and their optimization problems are summarized in table \ref{table2} and are concluded as follows:
\begin{enumerate}
\item Access Class Barring (ACB) Scheme: devices are forbidden to transmit preamble according to a probability $P_\text{ACB}$ chosen by BS to alleviate network congestion \cite{wu2012fast,he2015traffic,duan2016d,Luis2018Reinforcement}. 
\item Dynamic Resource Allocation (DRA) Scheme: BS allocates a number of channels for RACH according to the requirements during congestion \cite{sharma2019collaborative}. 
\item Back-off (BO) Scheme: different BO timers are assigned to different service classes in order to postpone their access attempts \cite{jian2013novel}. 
\item Prioritized Access Scheme: devices are splitting into several classes, where the devices from one class are allowed to perform access only in the dedicated access cycle \cite{lin2014prada}. 
\item Distributed Queuing (DQ) Scheme: devices perform access based on a tree splitting algorithm to resolve the collisions by organizing the re-transmission of colliding devices into several distributed queues \cite{vazquez2017contention}. 
\end{enumerate}

\section{Conventional Random Access Optimization}

Despite that each scheme introduced in Sec. \ref{sec3a} has its own mechanism to control access overload, these access schemes are intrinsically based on f-ALOHA protocol, which formulates a general discrete time stochastic control process. In detail, each scheme would divide time into frames, and allows a limited number of devices to execute access using a limited number of channels in each frame. The BS organize devices' transmission and re-transmission in a centralized manner to facilitate overload control in various traffic scenarios. Taking the ACB scheme for example, the BS controls the probability of device access by using the ACB factor, and each non-empty device randomly decides whether to execute RACH according to the obtained probability.
\color{black}

\vspace*{-0.2cm}
\subsection{Research Challenges of Random Access optimization}

RACH optimization targets to identify the optimal strategy to select RACH control parameters in real-time to optimize one or more KPIs. This optimal strategy of each RACH scheme is determined by an agent at the BS, which makes decision according to the received observations. More precisely, the observation is a set of historical transmission receptions during the RACH, including, but not limited to, the numbers of channels' state in success/collision/idle at the end of each frame, and the output of the agent is a set of parameters that will be performed in the forthcoming frame to maximize the KPIs in the following frames. Note that obtaining an optimal RACH configuration strategy using Bayesian approach is generally intractable. To tackle these problems, the optimization can be divided into two strongly related sub-tasks, including \emph{(a)} traffic load prediction for the forthcoming frame; and \emph{(b)} RACH control configuration based on the predicted traffic load. Taking the adaptive ACB scheme as an example, by predicting the forthcoming traffic statistic $\hat{N}$, the number of access success devices can be optimized by choosing the ACB factor $P_\text{ACB}$ according to $P_\text{ACB}=\text{min}(1,\frac{R}{\hat{N}})$, where $R$ refers to the number of channels.

\subsubsection{Traffic Prediction}
Traffic prediction can be the most critical problem in f-ALOHA based network, due to the following three challenges: \emph{(i)} the incoming traffic is generally complex, and may consist of mixtures of different traffic types including periodic, event-driven (bursty), multimedia streaming patterns, and etc.; \emph{(ii)} the lack of information about the cardinality of collisions; and \emph{(iii)} with overload network, transmission might be restricted according to the mechanism of each RACH scheme, which can lead to temporal correlations of traffic due to the unobserved packets accumulation.

\subsubsection{RACH Control Configuration}\label{sec3b}

Even with known predicted traffic statistics, maximizing the long-term KPIs for RACH is generally mathematical intractable, due to that these KPIs are not only determined by the current configuration, but also correlated with the future configurations. Most non-ML works only optimize the KPI of the next frame, where they ignored the dependency among the RACH control configurations of multiple consecutive frames over the long-term KPI. This simplified assumption of traffic is made due to the limitation in mathematical tool to capture these complex long-term correlation over traffic and the RACH control configurations. For the schemes introduced in Sec. \ref{sec3a} with the aim of optimizing the number of success access devices, the ACB scheme \cite{duan2016d} and the resource allocation scheme \cite{sharma2019collaborative} offered exact closed-form solutions, while the back-off scheme \cite{jian2013novel}, the prioritized access scheme \cite{lin2014prada} and the distributed queuing scheme \cite{vazquez2017contention} only enjoyed approximated solutions.

\vspace*{-0.2cm}
\subsection{Conventional non-ML based Access Control Optimization}\label{sec4a}

Given a known RACH control configuration strategy, the traffic prediction problem can be cast as a Bayesian probability inference problem, requiring the calculation of the probability for each possible traffic statistics under given historical observation. However, due to the lack of a priori probabilistic model for traffic generation, it is impossible to compute the probability of each occurring status. To solve this problem, previous works \cite{wu2012fast,he2017traffic,duan2016d} assumed that the traffic load statistic to be predicted at the $(t+1)$th frame (forthcoming) is equal to the traffic load at the frame $t$ (current). Note that this assumption is necessary to enable tractability of traffic prediction in non-ML based methods. The details of existing traffic estimation are concluded as:

\begin{enumerate}
\item Drift Analysis (DA) estimator: A unified framework for traffic load estimation was proposed in \cite{wu2012fast}, where the estimation was updated via a heuristic recursion algorithm. The traffic value is calculated by summarizing the last traffic value as well as the estimated traffic difference between the last and the current frames. The traffic difference estimation determines the prediction accuracy relying on the selection of adjusting parameters, which may be determined by either network analysis or trial-and-error processes.

\item Method of Moment (MoM) estimator: Given a specific traffic load value, the expected numbers of idle, success, and collision channels can be calculated using \cite[Eq. (27)]{duan2016d}. MoM predictor aims at matching one or more of the moments (i.e., the expectations) to the current observations, respectively. In other word, the MoM estimator determines the traffic load value that minimizes the discrepancy between the moments and its respective observations. Simplified MoM estimators have been proposed in \cite{duan2016d,jiang2019deep}, which enjoy closed-form solutions, with generally lower prediction accuracy. For instance, in \cite[Eq. (17)]{jiang2019deep}, traffic load was estimated by matching the first moment of idle preambles with the current observation of the number of idle preambles. 

\item Maximum Likelihood Estimator (MLE): MLE calculates the maximum likelihood of the optimal Bayes estimator with respect to each traffic load value under each given current observation. This is done in \cite{he2017traffic} by assuming that, in a frame, devices sequentially and independently select channels one after another, rather than selecting channels simultaneously. This sequential channel selection can be represented by a Markov chain, where the maximum likelihoods for each traffic load statistics of all observations can be calculated using the steady-state probability vector of the Markov chain.

\end{enumerate}

\section{Learning-Based Access Control Optimization}\label{sec4}

According to the high complexity of access control optimization, ML is a potential tool to provide better optimization performance than conventional methods. Due to the milliseconds level requirement of the RACH response time \cite{LTE2013dahlman,sharma2019towards}, it highly fit the ML methods with decoupled feedforwarding and training (e.g., DQN). In this scenario, the ML agent only feedforwards the ML agent, whose response time is acceptable, and this procedure only consumes little computational resource. Furthermore, the ML agent in the BS can be replaced by an updated one in anytime, where the updating can occur in an edge or a cloud, training by the collected data from all BSs. To do so, in the following, we first introduce conventional one-step RL-based access control optimization, and then propose a two-step learning based access control optimization, which decouples traffic prediction and parameter configuration.
\color{black}

\vspace*{-0.2cm}
\subsection{One-step Reinforcement Learning Based Access Control Optimization}\label{sec4b}

The non-ML based RACH access control optimization in Sec. \ref{sec4a} produces explicit optimization instructions, where its performance is limited due to the strong reliance on a mathematical model that captures the regularities of the practical communication environment. Instead of using explicit instructions, ML-based access control optimization expects to perform the RACH access control optimization relying on patterns and inference. These patterns and inference are obtained by training a “machine”, also known as a hypothesis class, to discover regularities in data using computational approach, rather than acquiring domain knowledge via the constructed physics-based model. Generally, by choosing a powerful “machine”, the ML-based method can better understand the environment than the constructed mathematical model \cite{gacanin2019artificial,sharma2019towards}.

The access control optimization can be formulated as a general Partially Observed Markov Decision Process (POMDP, a.k.a., belief MDP) problem. In detail, at the beginning of each frame, the BS assesses the traffic overload condition of the network relying on the observation of the network environment; then, it determines the RACH control action according to the observation using an optimization algorithm (eg., convex optimization, ML); finally, the BS evaluates the executed RACH control action at the end of each frame. The partial observation here refers to that the BS is unable to know all the information of the communication environment, including, but not limited to, the random collision and the traffic generation processes.
\color{black}

The RL algorithms are well-known in addressing dynamic control problem in complex POMDPs by learning the optimal control strategy via interacting with environment \cite{sutton2017reinforcement}. In access control optimization, the POMDP problem (in Sec. \ref{sec3b}) can be described as a four-tuple $\{$\emph{state, action, reward, new state}$\}$, and an RL-agent is located at the BS to discover the optimal RACH control configurations in each frame yielding the best long-term KPI by trying them. To memorize these experiences, the RL-agent parameterizes the action-state value function that be represented by either an exact table, or an approximated function (e.g., linear function, neural network, decision tree, and etc.). The training progress is briefly concluded as: 1) at the beginning of each frame, the RL-agent configure RACH parameters (\emph{action}) by feeding the unprocessed observation (\emph{state}) into the value function; 2) devices initiate access according to the RACH control parameters; and 3) the RL-agent receives a scalar (\emph{reward}) to evaluate the current access performance, and updates its value function using Bellman equation.

The RL algorithms have proven to be useful in several applications in the area of access control optimization \cite{Luis2018Reinforcement,sharma2019collaborative,jiang2019deep}. Recent work \cite{Luis2018Reinforcement,sharma2019collaborative} has proposed tabular Q-learning based congestion minimization schemes, which aim at selecting optimal (\emph{action}) to maximize the number of success accesses (\emph{reward}). These Q-learning based schemes are shown to outperform the conventional dynamic ACB scheme given in \cite{duan2016d}. Unfortunately, a direct application of the tabular RL algorithms may not be feasible to solve all the access optimization problems. For instance, tabular RL algorithms cannot be adopted in NB-IoT networks, due to that its large size of the action and state space results in low training efficiency. To solve it, Deep Reinforcement Learning (DRL) was adopted to enable learning over a large state space in \cite{jiang2019deep} inspired by intelligent game playing \cite{sutton2017reinforcement}, while the action space was broke down into several action variables to be cooperatively trained by multiple agents that solves the oversize action space problem \cite{jiang2019deep}.

\vspace*{-0.2cm}
\subsection{Supervised Learning Based Traffic Prediction with Conventional RACH Control Configuration}\label{secVa}
\vspace*{-0.0cm}

Adopting one-step DRL based access control optimization methods proposed in Section \ref{sec4b} still face several challenges including:
\emph{a)} training DRL agent requires huge computational resource; 
\emph{b)} the DRL agent is less interpretable and reliable due to the basis of “black box” characteristics;
and \emph{c)} the DRL agent is expected to be updated in an online manner, but the convergence is really slow due to the complexity of value function as well as the tradeoff between exploration and exploitation. To solve these problems, we propose two-step ML-based optimization methods based on individual learning in the traffic prediction as well as in the RACH control configuration in follows.

Given a Non-ML based RACH control configuration strategy, we can focus on solving the traffic prediction using learning-based method to improve the access performance, namely, SL-based optimizer. This learning-based predictor has been given in \cite{jiang2019online}, which applies a powerful “machines” to learn the mathematical models during traffic generation and communication. In Table \ref{table2}, we summarize the basic characteristics, performance, and efficiency of each access control optimization method.

This SL-based traffic predictor adopts a modern RNN model based on the Long Short-Term Memory (LSTM) architecture, which can capture traffic statistics over consecutive frames. The input of predictor is a set containing the historical RACH control configurations as well as the observation on the number of idle, collided, and successful channels in each frame. Different from those conventional methods only utilizing the most recent observation, this SL-based predictor captures the historical information from the previous observations to learn the time varying traffic trend for better prediction accuracy.

The LSTM RNN is trained by leveraging a novel approximate labeling technique that is inspired by MoM estimators given in Sec. \ref{sec4}. This approximate labeling technique enables online training in the absence of feedback on the exact cardinality of collisions. This online adaptation allows LSTM RNN to adapt to the traffic statistics in runtime. In details, LSTM RNN is progressively fed with a finite observations to produce the predicted traffic at beginning of frame $t$. At frame $t+1$, an error-correction traffic value can be estimated using the exact transmission receptions over the frame $t$ using any traffic estimation method described in Sec. \ref{sec4a}. In this way, with one frame delay, the weights of the LSTM RNN can be adjusted in order to minimize the error of the traffic value predicted by the LSTM RNN at frame $t$ with respect to the error-correction traffic value estimated at frame $t+1$.

\vspace*{-0.0cm}
\captionsetup{singlelinecheck=false} 
\begin{figure}[htbp!]
    \begin{center}
        \includegraphics[width=0.45\textwidth]{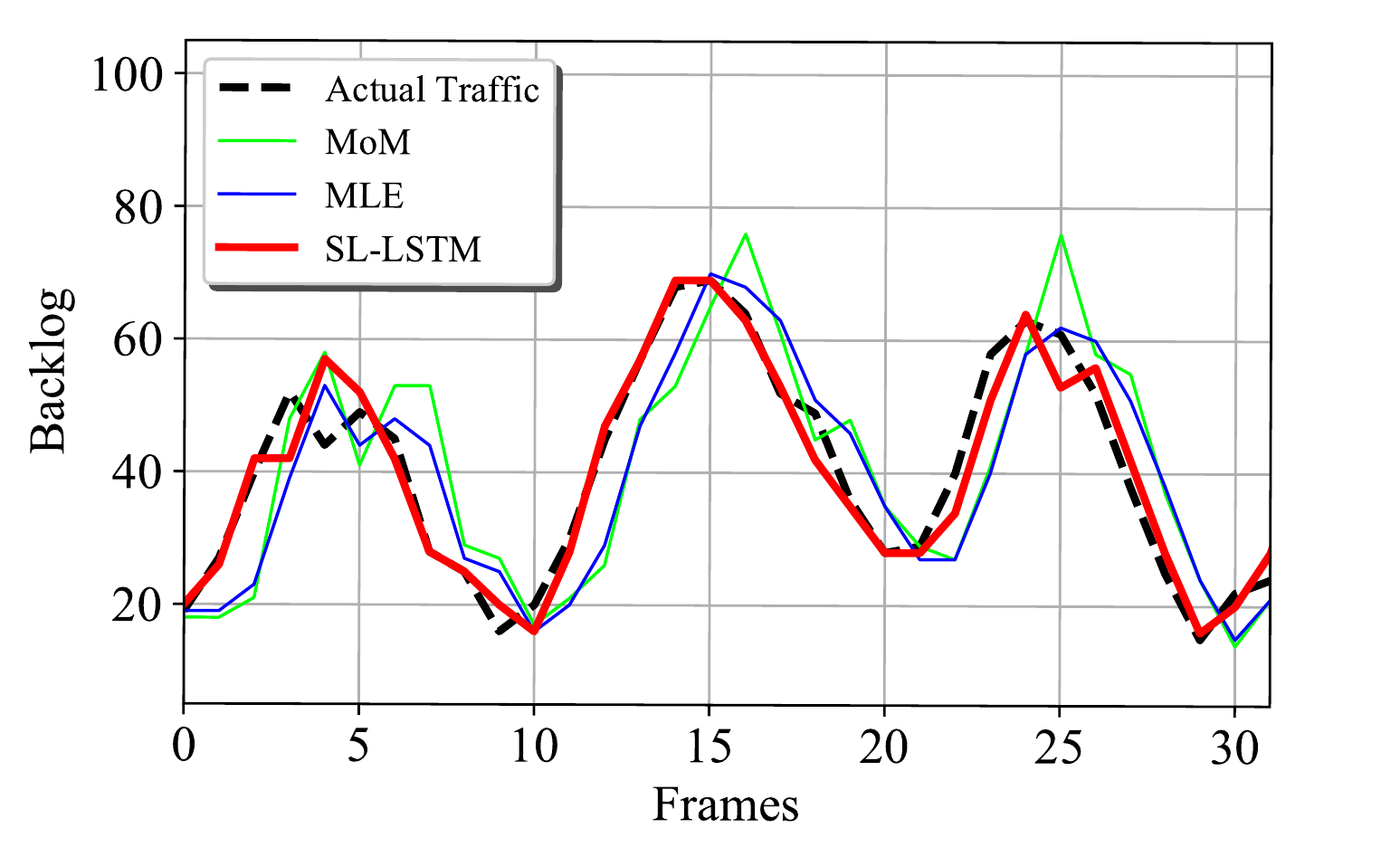}
         \caption{The actual and predicted backlog of each predictor.}
        \vspace*{-0.3cm}
         \label{fig3}
    \end{center}
    \vspace*{-0.2cm}
\end{figure}

\vspace*{-0.0cm}
\captionsetup{singlelinecheck=false} 
\begin{figure}[htbp!]
    \begin{center}
        \includegraphics[width=0.45\textwidth]{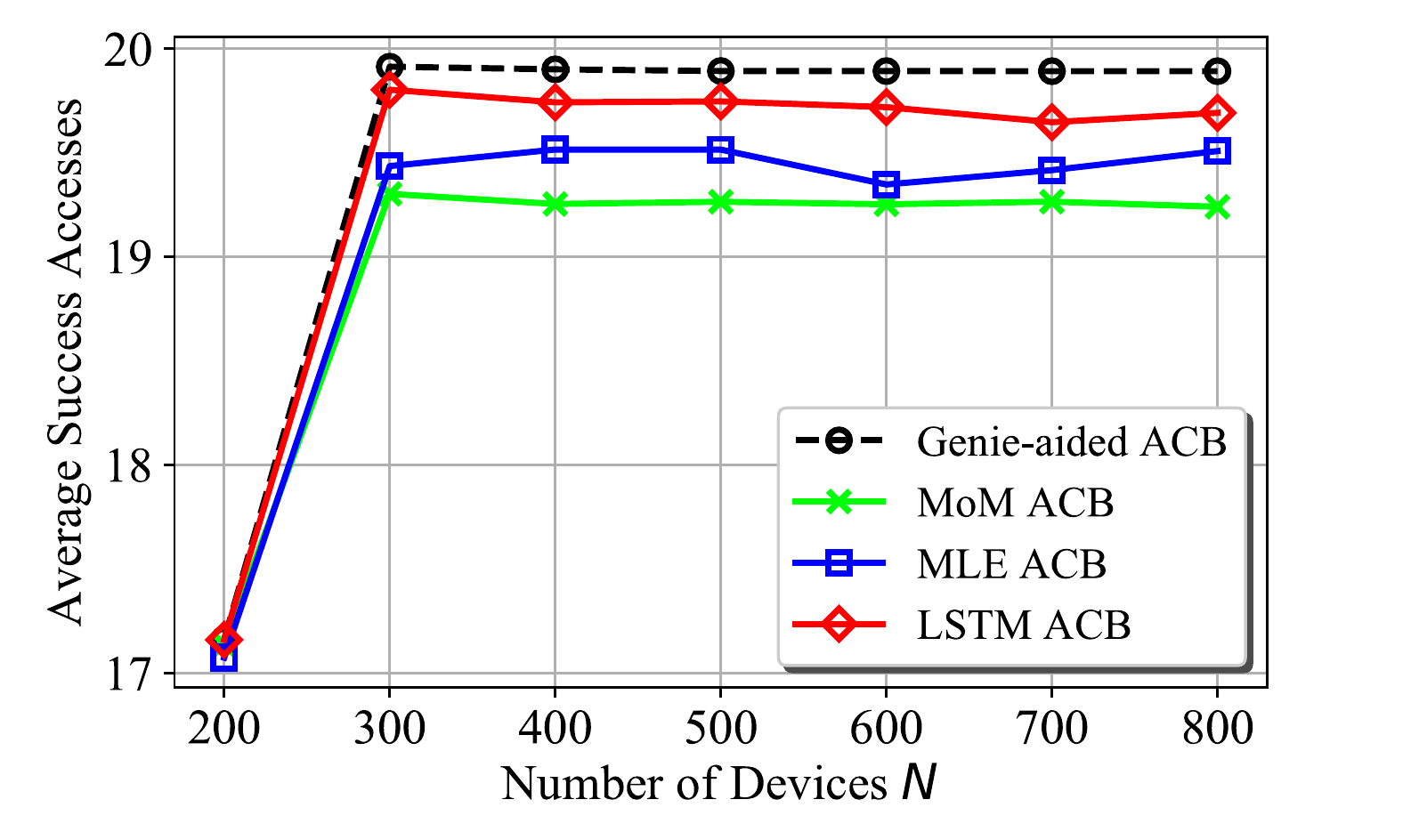}    
        \caption{The average number of success access devices per episode of each optimizer in the ACB scheme.}
        \vspace*{-0.3cm}
                \label{fig5}
    \end{center}
    \vspace*{-0.2cm}
\end{figure}

\vspace*{-0.0cm}
\captionsetup{singlelinecheck=false} 
\begin{figure*}[htbp!]
    \begin{center}
        \begin{minipage}[t]{1\textwidth}
    \centering
        \includegraphics[width=1\textwidth]{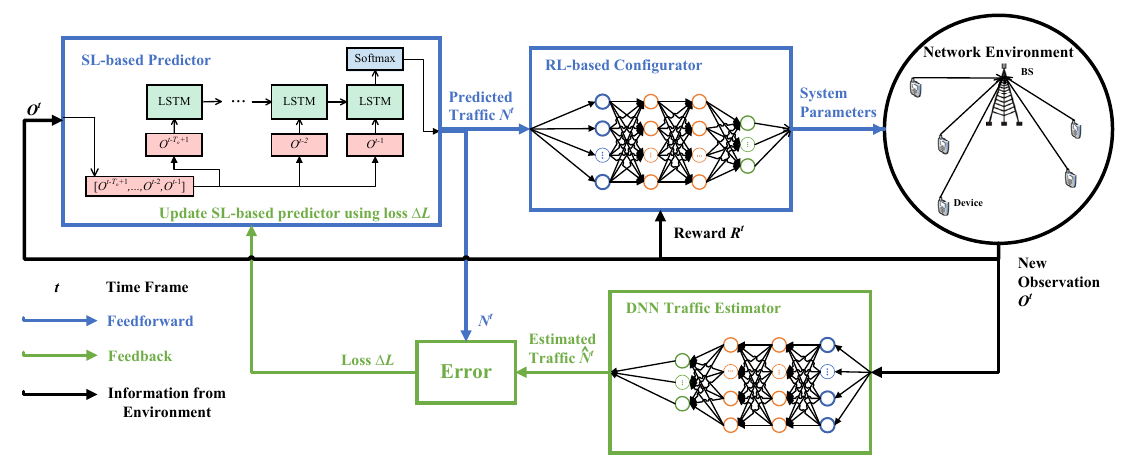}
        \vspace*{-0.7cm}
        \caption{\scriptsize Illustration of the feedforward and the online adaptation of the multi-step CPCL optimizer.}
                \label{figCPCL}
        \end{minipage}
    \end{center}
    \vspace*{-0.5cm}
\end{figure*}

The results are obtained by simulations using Python by comparing the traffic prediction accuracy and the average number of success access devices of the ACB scheme with the MoM optimizer, the ML optimizer, and the SL-based optimizer using LSTM RNN. In simulations, we set the number of channels as $54$, the retransmission constraint as $10$, and the traffic as the time limited Beta profile with parameters $(3,4)$ repeated every $10$ frames (The following results in Sec. V.B are also based on these network parameters). Fig. \ref{fig3} plots the actual and predicted backlog of each predictor, where the SL-based result is obtained by the predictor trained over $10^5$ frames. We observe that only SL-LSTM can predict the backlog spikes coming from bursty traffic, due to their capability in capturing historical trends of time-varied traffic. Fig. \ref{fig5} plots the average number of success access devices per episode (each containing 100 frames) of each optimizer and the “Genie-aided ACB” (i.e., referring to the ACB scheme aided by actual backlog). It is seen that the SL-based optimizer outperforms the other optimizers, due to its better prediction accuracy. However, it should be emphasized that each optimizer relies on the exact ACB configuration solution. Once the RACH scheme becomes complex (e.g., the hybrid ACB and Back-Off (ACB$\&$BO) scheme and the DQ scheme), the access performance may be degraded due to the ineffectiveness of non-ML based access control configuration.

\vspace*{-0.2cm}
\subsection{Supervised Learning Based Traffic Prediction with Reinforcement Learning Based Access Control Configuration}\label{secVb}
The SL-based optimizer relies on non-ML RACH controller. To optimally perform the traffic prediction as well as access control, in this subsection, we propose a two-step Cooperative Prediction and Control Learning (CPCL) optimizer, which individually executes the RNN traffic prediction and the DRL-based access control configuration as shown in Fig. \ref{figCPCL} using the following 4 steps:
\begin{enumerate}
\item At the beginning of any frame $t$, the RNN predictor is to produce the traffic load $N^t$ using the raw observations $[O^{t-T_o+1},\cdots,O^{t-2},O^{t-1}]$ as Sec. \ref{secVa}; 
\color{black}
\item The predicted traffic load is input into the DRL-agent to control random access parameters;
\item At the next frame $t+1$, the error-correction traffic value $\hat{N}^t$ can be estimated by a fully connected Deep Neural Network (DNN) using the new observation $O^{t}$;
\item Finally, the SL-based predictor is updated by minimizing the error $\Lambda L$ of the predicted traffic value $N^t$ via the SL-based predictor at frame $t$ with respect to the error-correction traffic value $\hat{N}^t$ estimated via DNN estimator at frame $t+1$.
\end{enumerate}


The use of DRL-based configuration in lieu of the conventional parameters configuration in Sec.\ref{secVa} enables more effective configuration in high complexity access schemes. Note that the RNN traffic predictor and the DRL-agent for access control configuration in CPCL-based optimizer can be updated at runtime, which allows it to adapt to the realistic traffic. Conversely, the DNN-agent for error-correction traffic estimation can be trained using the offline rather than online, due to its independence on the real dynamic traffic in the environment. Compared to the one-step DRL-based optimizer, this two-step CPCL optimizer may take much less training time with less computational resource. Furthermore, the two-step CPCL optimizer can achieve optimization for multiple access control parameters configuration using multiple agents but sharing only one traffic predictor, which can ease the implementation.


\vspace*{-0.0cm}
\captionsetup{singlelinecheck=false} 
\begin{figure}[htbp!]
    \begin{center}
        \includegraphics[width=0.45\textwidth]{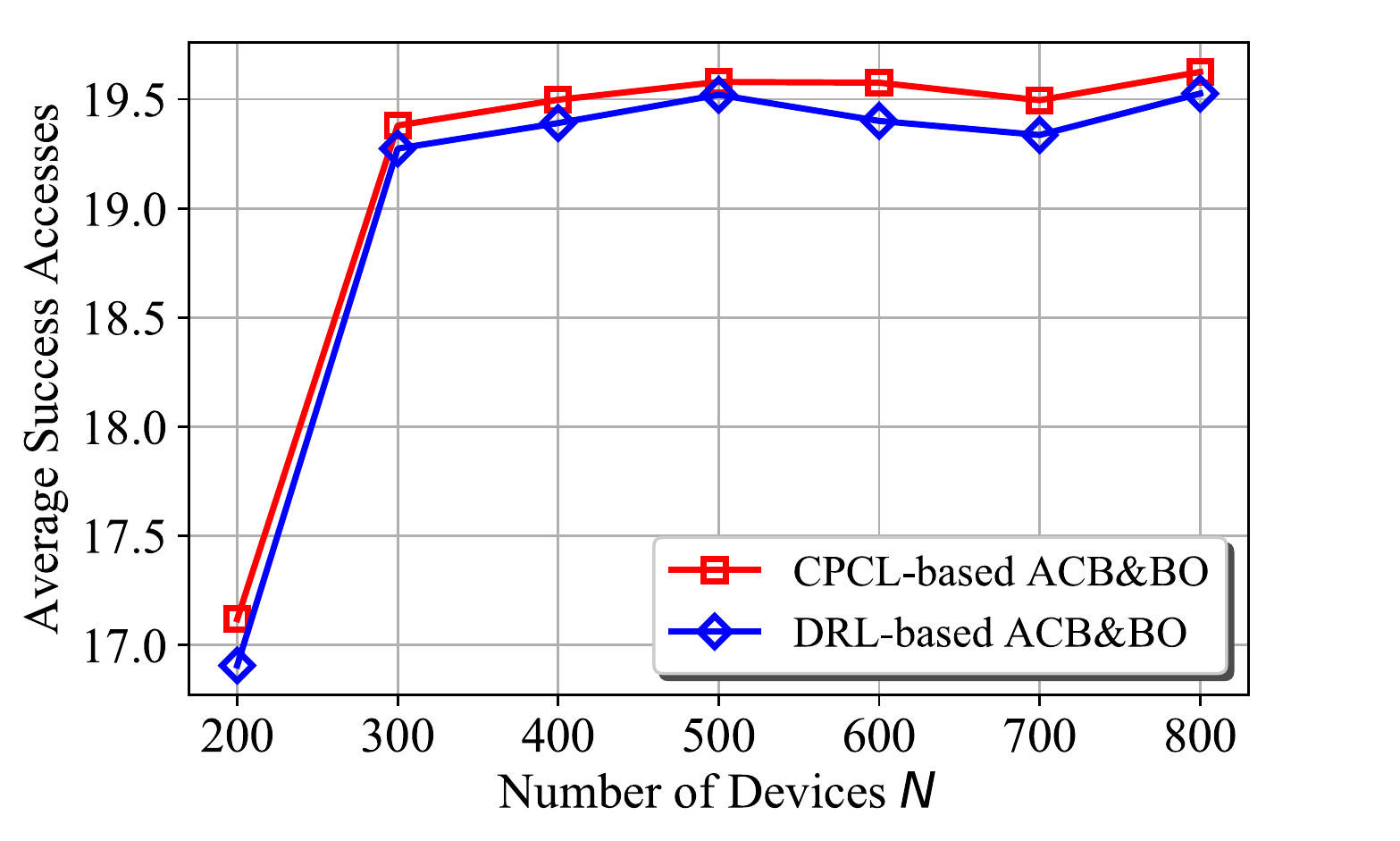}
        \caption{The average number of success access devices per episode of DRL-based optimizer and CPCL-based optimizer in the ACB$\&$BO scheme.}
        \vspace*{-0.5cm}
                \label{fig6}
    \end{center}
    \vspace*{-0.0cm}
\end{figure}

Fig. \ref{fig6} compare the average number of success access devices per episode of the DRL-based optimizer and the CPCL-based optimizer for the ACB scheme and the ACB$\&$BO scheme, respectively. Both figures shown that the CPCL-based optimizer slightly outperforms the DRL-based optimizer due to that the CPCL-based optimizer is capable of converging to a better solution. The advantage of the CPCL-based optimizer may come from the fact that it breaks down the optimization procedure into two sub-learning problems, which makes learning easier.

\begin{figure}[htbp!]
    \begin{center}
        \includegraphics[width=0.45\textwidth]{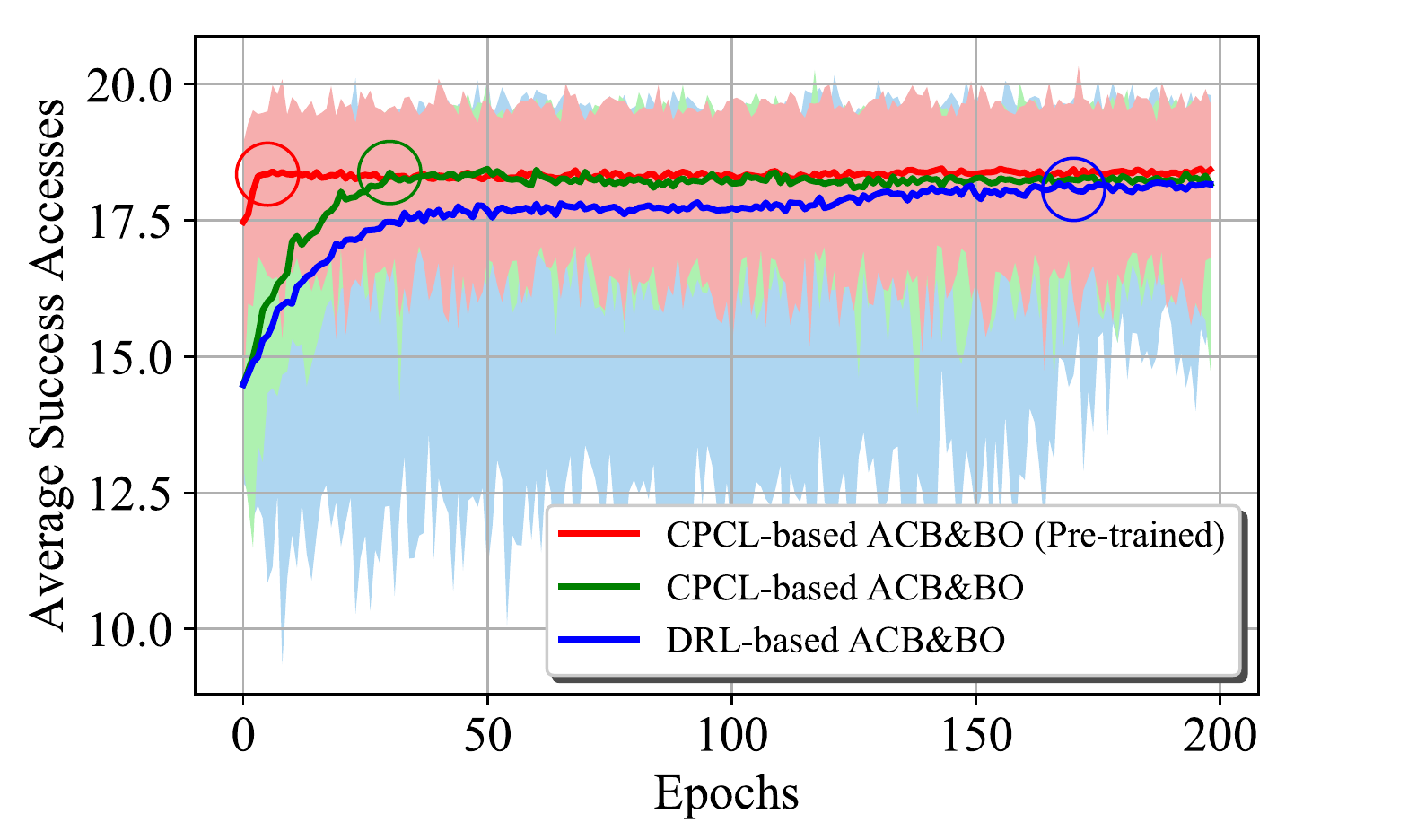}
        \vspace*{-0.3cm}
        \caption{The required number of episodes for each ML-based optimizer that converges to an efficient solution, where each optimizer has the same hidden layers of neural network and the same hyperparameters for training.}
                \label{fig7}
    \end{center}
    \vspace*{-0.5cm}
\end{figure}

Fig. \ref{fig7} plots the evolution (averaged over 200 training trails) of the average success accesses per frame as a function in the online phase for ``CPCL-based ACB$\&$BO (Pre-trained)", ``CPCL-based ACB$\&$BO (Pre-trained)", and ``DRL-based ACB$\&$BO". Here, the ``CPCL-based ACB$\&$BO (Pre-trained)" refers to that its DRL-agent for access control configuration has been pre-trained, while ``CPCL-based ACB$\&$BO" is without any pre-training. The approximated converging point of each scheme is highlighted by circles. It is seen that the pre-training can help the CPCL-based ACB$\&$BO optimizer to be fairly faster to converge than it without pre-training. It can also be observed that the training speed of ``CPCL-based ACB$\&$BO (Pre-trained)" (consumes about 2 episodes) is about 100 times faster than the DRL-based optimizer (consumes about 170 episodes), which sheds light on its capability of its efficient adaptation. Furthermore, CPCL-based optimizers show better performance than the DRL-based ACB$\&$BO after convergence.
\color{black}



\vspace*{-0.0cm}
\captionsetup{singlelinecheck=false} 

\section{Conclusion and Future Work}
\vspace*{-0.0cm}


In this article, we elaborated ML techniques to be applied in access control optimization for random access schemes, which has the potential to play an essential role in realizing the efficient access of future wireless networks. The conventional single-step DRL-based optimizer is shown to outperform the non-ML based optimizers in terms of success access devices, due to that it is capable of learning to master the challenging optimization task. However, the single-step DRL-based optimizer suffers from low training efficiency and the requirement of huge computational resource. To solve this problem, we proposed two-step CPCL-based optimization methods to individually learn the traffic prediction and the RACH control configuration, which considerably improved the training efficiency.

Our results revealed that ML techniques have great potential to revolutionize access control optimization. Compared with the conventional DRL-based method, the proposed CPCL-based method can achieve higher training efficiency and better access performance, and can be applied for performance optimization of other types of random access schemes. Furthermore, we have identified following future research directions: 1) develop transfer learning and meta-learning for online updating to improve training efficiency; 2) develop distributed learning at devices and BSs to cooperatively guide the transmission decisions; and 3) exploit learning based priority-aware optimization for heterogeneous applications.



\appendices
\numberwithin{equation}{section}

\ifCLASSOPTIONcaptionsoff
  \newpage
\fi

\bibliographystyle{IEEEtran}
\bibliography{IEEEabrv,RA_bib}

\vskip -2\baselineskip plus -1fil

\begin{IEEEbiographynophoto}{Nan Jiang} is currently working toward the Ph.D. degree in electronic engineering at Queen Mary University of London, London, U.K. His research interests include machine learning and internet of things.
\end{IEEEbiographynophoto}

\vskip -2\baselineskip plus -1fil

\begin{IEEEbiographynophoto}{Yansha Deng} 
is currently a Lecturer (Assistant Professor) with the Department of Informatics, King’s College London. Her research interests include internet of things, 5G wireless networks, and molecular communication. 
\end{IEEEbiographynophoto}

\vskip -2\baselineskip plus -1fil

\begin{IEEEbiographynophoto}{Arumugam Nallanathan} is Professor of Wireless Communications and Head of the Communication Systems Research (CSR) group in the School of Electronic Engineering and Computer Science at Queen Mary University of London. His research interests include Beyond 5G Wireless Networks, Internet of Things, and Molecular Communications. 

\end{IEEEbiographynophoto}

\end{document}